\documentclass[conference]{IEEEtran}
\IEEEoverridecommandlockouts
\usepackage{cite}
\usepackage{amsmath,amssymb,amsfonts}
\usepackage{algorithmic}
\usepackage{graphicx}
\usepackage{textcomp}
\usepackage{array}
\usepackage{multirow}
\usepackage{xcolor,mdframed}
\usepackage{float}
\usepackage{booktabs}
\usepackage{grffile}
\usepackage{authblk}
\newcommand{\comment}[1]{}
\def\BibTeX{{\rm B\kern-.05em{\sc i\kern-.025em b}\kern-.08em
    T\kern-.1667em\lower.7ex\hbox{E}\kern-.125emX}}

\newcommand{\slicex}{6}
\newcommand{\slicey}{5}
\newcommand{\recon}{ncxpdnet}

\begin{document}

\title{Learning the sampling density in 2D SPARKLING MRI acquisition for optimized image reconstruction}
%\title{2D SPARKLING acquisition with optimized target sampling density for MR image reconstruction network}

\author[1,2]{Chaithya G R\thanks{Chaithya G R was supported by the CEA NUMERICS program, which has received funding from the European Union's Horizon 2020 research and innovation program under the Marie Sklodowska-Curie grant agreement No 800945.
}}
\author[1,2,3]{Zaccharie RAMZI}
\author[1,2]{Philippe CIUCIU}%~\IEEEmembership{Senior Member, IEEE}
\affil[1]{CEA, Joliot, NeuroSpin, Universit\'e Paris-Saclay, F-91191 Gif-sur-Yvette, France}
\affil[2]{Inria, Parietal, Universit\'e Paris-Saclay, F-91120 Palaiseau, France}
\affil[3]{AIM, CEA, CNRS, Universit\'e Paris-Saclay, Universit\'e Paris Diderot, Sorbonne Paris Cit\'e, F-91191 Gif-sur-Yvette}

\maketitle

\begin{abstract}
The SPARKLING algorithm was originally developed for accelerated 2D magnetic resonance imaging~(MRI) in the compressed sensing~(CS) context. It yields non-Cartesian sampling trajectories that jointly fulfill a target sampling density while each individual trajectory complies with MR hardware constraints. However, the two main limitations of SPARKLING are first that the optimal target sampling density is unknown and thus a user-defined parameter and second that this sampling pattern generation remains disconnected from MR image reconstruction thus from the optimization of image quality. Recently, data-driven learning schemes such as LOUPE have been proposed to learn a discrete sampling pattern, by jointly optimizing the whole pipeline from data acquisition to image reconstruction. In this work, we merge these methods with a state-of-the-art deep neural network for image reconstruction, called XPDNet, to learn the optimal target sampling density. Next, this density is used as input parameter to SPARKLING to obtain 20x accelerated non-Cartesian trajectories. These trajectories are tested on retrospective compressed sensing~(CS) studies and show superior performance in terms of image quality with both deep learning~(DL) and conventional CS reconstruction schemes.
\end{abstract}


\section{Introduction}
Compressed sensing~(CS) in MRI~\cite{Lustig2007}  has led to a large reduction in scan time while maintaining a reasonable reconstructed MR image quality. Practically, CS is implemented by undersampling pseudo-randomly the k-space according to a variable sampling density~\cite{puy2011variable,chauffert2013variable, Chauffert_SIAM2014,adcock2017breaking,boyer2017compressed}.
The sampling pattern may be composed of multiple individual Cartesian lines~(Cartesian Sampling), in which case variable density sampling~(VDS) is implemented only along the phase encoding dimension. To go to higher reduction in scan times, non-Cartesian sampling is really helpful as it permits the implementation of 2D VDS with the help of non-Cartesian trajectories, such as radial spokes~\cite{radial_spokes} and spiral interleaves~\cite{ahn1986high}). Although radial and spiral sampling are widespread, they are not really optimal as radial spokes don't cover the k-space perfectly and spiral interleaves do not exactly match a prescribed sampling density. Hence, severe artifacts impede image quality during CS reconstruction.

The Spreading  Projection  Algorithm  for  Rapid  K-space  samplING, or SPARKLING~\cite{Lazarus_MRM_19} has been introduced as an iterative scheme that optimizes for each k-space trajectory to be compliant with MRI hardware constraints~(particularly maximum gradient and slew rate constraints), while ensuring that the overall sampling pattern obtained with all the trajectories follows a target sampling density. Further, the algorithm ensures that  optimized k-space sampling pattern does not have any local clusters, leading to locally uniform sampling patterns. This algorithm was extended to 3D \cite{Lazarus_NMRB_20,ChaithyaGR_IEEE_20} and showed superior performance in both terms of a peaky point spread function and image quality. 

However, a major drawback of SPARKLING algorithm is the need to setup a target sampling density as an input to the algorithm. In our earlier studies, we relied on heuristic methods to set this sampling density. The latter was parameterized to be radially decaying and its optimal parameters~(decay, cutoff) were grid searched during retrospective reconstruction studies in which image quality was maximized as a function of optimized trajectories for varied target densities.
%The sampling pattern was parameterized based to be a radially decaying densities, and the optimal parameters were grid searched based on retrospective reconstruction performance of the optimized trajectories for varied target densities.
However, this approach is too computationally expensive. Also, with a parameterized target density, the search space is too constrained, preventing us to obtain organ, imaging-contrast or orientation-specific sampling schemes. %or imaging contrast  
% This prevents us from obtaining more complex sampling patterns which could be specific to a given organ, orientation or imaging contrast.
One way to tackle this problem is by learning the target sampling density using data-driven approaches.

In~\cite{Knoll2011}, the authors proposed a naive approach to choose the target sampling density by averaging the power spectra of multiple MR images in a dataset. This method results in sampling densities that enforce denser sampling in the low frequencies. %where most of the power is concentrated in the k-space, particularly 
In~\cite{Knoll2011} the authors showed that this approach outperforms standard VDS and remains robust to variability in anatomy and orientation. However, this method focuses purely on the MRI dataset and is agnostic to the reconstruction technique. All MRI reconstruction algorithms enforce a prior (like sparsity in the wavelet or image gradient domain). Recent deep learning~(DL) reconstruction algorithms~\cite{Knoll_2020,muckley2021results,ramzi_bench2020} have learned more complex priors based on the organ or contrast-specific dataset. The target sampling density can be more efficient if it takes these priors into account and enforce denser samples in regions where the degree of uncertainty associated with such priors for reconstruction is higher. % used

More recently, methods like~\cite{sherry2020learning, loupe2020} learn the sampling pattern for MRI in a data-driven manner while optimizing for image quality at the reconstruction stage.  %In~\cite{sherry2020learning}, the authors present a novel bi-level optimization algorithm, which learns the sampling pattern for a convex CS based reconstruction.
In the deep learning setting, LOUPE~\cite{loupe2020} jointly optimizes the sampling density and the weights of a U-net architecture for image reconstruction. However, these studies are limited to Cartesian sampling. 
%setting and present their results with retrospective studies while using a binary mask obtained by sampling according to the target density.
Most appealing contributions~\cite{weiss2019pilot, vedula20203d} tend to directly learn the trajectories in a data-driven manner under MR hardware constraints. Particularly, in \cite{weiss2019pilot}, the authors use multi-resolution to overcome the problem of a large number of trainable parameters which crops up in such direct optimization. However, the final trajectories were similar to perturbed versions of the initialization.

In this work, we use the target density obtained by LOUPE as an input to the SPARKLING algorithm to generate 2D SPARKLING non-Cartesian trajectories. We carry out retrospective studies and compare them with those that result from other densities such as the average (log-)power spectra over the fastMRI dataset. We perform image reconstruction using both CS technique and the newly developed NC-PDNet~\cite{ramzi2021density} architecture which is a density compensated unrolled neural network for non-Cartesian MRI reconstruction. We conclude that the proposed solution~(LOUPE+2D~SPARKLING) outperforms other VDS approaches in terms of image quality.

\section{Materials and methods}

Here, we detail the methods used to optimize for sampling density and thereby how the latter is injected as an input to SPARKLING algorithm to generate non-Cartesian trajectories. We later study the performance of the corresponding sampling schemes on retrospective MR image reconstruction studies.

\subsection{2D Non-Cartesian trajectories}
Throughout this work, we follow the formulation we developed in~\cite{ChaithyaGR_IEEE_20}, for the case of 2D imaging. Let the MR image size be $N \times N$, over a field of view $\mathcal{F} \times \mathcal{F}$. Then the 2D k-space of the image is defined in $[-K_{\rm max}, K_{\rm max}]^2$, with $K_{\rm max} = \frac{N}{2 \mathcal{F}}$. In all our trajectories, we kept $N=320$ and $\mathcal{F}=0.23$ m. For the sake of simplicity, let us normalize the k-space to $\Omega = [-1, 1]^2$.  % \times [-K_{\rm max}, K_{\rm max}]
We are optimizing for the 2D k-space sampling pattern $\mathbf{K}$ which is composed of several shots $N_c$, $\mathbf{K}=(\mathbf{k}_i)_{i=1}^{N_c}$. Each 2D shot $\mathbf{k}_i(t)=(k_{i,x}(t), k_{i,y}(t))$ is controlled by the magnetic field gradients $\mathbf{G}_i(t)= (G_{i,x}(t), G_{i,y}(t))$ as follows: 
%\begin{align}\label{eq:ktraj}
$\mathbf{k}_i(t) = \frac{\gamma}{2\pi} \int_0^t \mathbf{G}_i(\tau)\,d\tau \,$
%\end{align}
with $\gamma$ the gyro-magnetic ratio~($\gamma=42.57$MHz/T for proton imaging). Each shot is sampled at the pace of gradient raster time $\Delta t$, throughout the readout time $T_{obs}$, resulting in $N_s = \lfloor \dfrac{T_{\rm obs}}{\Delta t}\rfloor$ samples per shot. The k-space data from the scanner is sampled at dwell time $\delta t$, which in practice is a fraction of $\Delta t$. Thus the total received k-space samples are of the form $\mathcal{K} \in \mathbb{C}^{N_c \times N_s \times \frac{\Delta t}{\delta t}}$. In our studies, we used dwell time ($\delta t=2\mu$s) and gradient raster time ($\Delta t=10\mu$s), thereby having 5 times more k-space sample points than the measurements defined by the gradient wave forms.

The MR hardware constraints of maximum gradient strength ($G_{\rm max} = 40$ mT/m) and slew rate ($S_{\rm max} =180$ T/m/s) results in a constrained trajectory with limited speed ($\alpha$) and acceleration ($\beta$). Note that the speed constraint also handles the Nyquist sampling criterion~(see \cite{chauffert2017projection}). We define this constraint set as $\mathcal{Q}_{\alpha, \beta }^{N_c}$, see~\cite{ChaithyaGR_IEEE_20}.

\subsection{SPARKLING algorithm}
Let the target sampling distribution be $\rho:\Omega\to \mathbb{R}$, with $\rho(x)\geq 0$ for all $x$ and $\int \rho(x) \,dx=1$. Given $\rho$, the SPARKLING algorithm optimizes for the k-space sampling pattern $\mathbf{K}$ such that the actual sampling distribution is closest to $\rho$, while being locally uniform. Although theoretically SPARKLING takes a continuous distribution $\rho$ as input parameter, in practice, we discretize the distribution to obtain $\boldsymbol{\rho} \in \mathbb{R}^{N \times N}$.
Further, the algorithm ensures that the each k-space shot $\mathbf{k}_i(t)$ in optimal $\mathbf{\widehat{K}}$ lies in $\mathcal{Q}_{\alpha, \beta }^{N_c}$. 
We can now summarize the SPARKLING algorithm as follows: 
\begin{equation}
	\mathbf{\widehat{K}} = \mathcal{S}(\boldsymbol{\rho}, \mathcal{Q}_{\alpha, \beta }^{N_c}, \mathbf{K_0})
\end{equation}
with $\mathbf{K_0}$ being the initialization. The detailed algorithm is presented in~\cite{ChaithyaGR_IEEE_20}.
Hereafter, we discuss different gridded distributions $\boldsymbol{\rho}$ that were obtained for our study.

\subsection{Target sampling density learning}
\label{sec:density_spec}
%One of the primary inputs to the SPARKLING algorithm is the target sampling distribution $\rho$. 
In this work, we broadly use four methods for estimating or learning a target sampling density. All these methods are data-driven and we rely on the fastMRI dataset~\cite{zbontar2019fastmri} to compute them. Let $\left\{\mathbf{x}_j \in \mathbb{R}^{N \times N}\right\}_{j=1}^n$ denote brain MR images from this dataset, where $j$ is the scan number and $n$ is the total number of images~(for simplicity, we used magnitude-only images). Let $\left\{\mathbf{v}_j \in \mathbb{C}^{N \times N}\right\}_{j=1}^n$ correspond to their respective discrete k-spaces on a grid~(Fourier spectrum) obtained by a fast Fourier transform.
%For the sake of comparison, we also obtained variable densities with conventional parameterization as a baseline for our study.

\subsubsection{VDS-based ($\boldsymbol{\rm \rho}_{vds}$)}
The first method we employed to obtain a density is based on naive VDS. For this, we parameterized the density as radially decaying with cutoff C and decay D as described in \cite{ChaithyaGR_IEEE_20}:
\begin{equation}
\label{eq:target_distribution}
\rho_{vds}^{C, D}(x) = 
\begin{cases}
\kappa &\quad|x| < C \\
\kappa \left(\frac{C}{|x|}\right)^D &\quad|x| > C
\end{cases}
\end{equation}
In our experiments, we heuristically grid searched for optimal parameters and used $C=25\%$ and $D=2$ as the best density.

\subsubsection{Spectrum-based ($\boldsymbol{\rm \rho}_{sb}$, $\boldsymbol{\rm \rho}_{lsb}$)}
Next we obtained a sampling density based on~\cite{Knoll2011} which involves averaging the spectra of brain images from the fastMRI dataset. Let $\mathbf{v}_{avg}$ correspond to the average of all the spectra $\mathbf{v}_j$ in the dataset. Then we can normalize the 2D spectrum to obtain a sampling density $\boldsymbol{\rho}_{\rm sb}$ on the $N \times N$ grid:
\begin{equation}
	\label{eq:normalize_pdf}
	\boldsymbol{\rho}_{\rm sb}(p,q) = \frac{\mathbf{v}_{avg}(p,q) - \min (\mathbf{v}_{avg})}{\sum_{p,q} \left[\mathbf{v}_{avg}(p,q) - \min (\mathbf{v}_{avg}) \right]} \, .
\end{equation}

Further, we observed that the spectra have very large magnitudes at lower frequencies as compared with higher frequencies. In an effort to flatten the distribution so that we may better balance all frequencies, we relied on an average {\em log-spectrum} $\mathbf{v}_{lavg}$ of the fastMRI images and obtained the distribution $\boldsymbol{\rho}_{\rm lsb}$ by replacing $\mathbf{v}_{avg}$ with $\mathbf{v}_{lavg}$ in Eq.~\eqref{eq:normalize_pdf}.

\subsubsection{LOUPE-based ($\boldsymbol{\rm \rho}_{lb}$)}
\label{sec:density_loupe}
As the spectrum-based methods are agnostic to image reconstruction, to fill this gap we used the Cartesian acquisition model from LOUPE~\cite{loupe2020}. LOUPE is actually a DL-based optimization scheme that learns a Cartesian under-sampling pattern for a prescribed sparsity level $\gamma$, which provides the percentage of  discarded measurements as compared to a full sampling. Hence, $\gamma$ is defined as the inverse of the under-sampling factor $R$ ($= \frac{N \times N }{N_c \times N_s \times \frac{\Delta t}{\delta t}} = \frac{1}{\gamma}$ for non-Cartesian sampling).
In practice, we used $R=2.5$~($\gamma=0.4$).
Using LOUPE, we can learn a gridded sampling density $\boldsymbol{\rho}_{lb}$ by jointly optimizing the acquisition and reconstruction frameworks in the Cartesian domain. In~\cite{loupe2020}, the authors used conventional U-Net~\cite{unet} for carrying out reconstruction. In contrast here, we integrate LOUPE's acquisition network with a modular cross-domain neural network called XPDNet~\cite{ramzi2020xpdnet} which stood second in the 2020 fastMRI brain reconstruction challenge~\cite{muckley2021results}. Hence, we jointly optimize for the sampling distribution $\boldsymbol{\rm \rho}_{lb}$ and the reconstruction network. In regards to the LOUPE model, we initialize the {\em sigmoid sample slope} $s=20$ and trained this network for 100 epochs over all the training set~($n=4469$ MR images) and probed for the target sampling density. We ensured that there was no leaking of the k-space data into the reconstruction network by checking the resulting binary sampling masks~(see~\cite{loupe2020} for details). % for \$T\_1\$-w training

\subsection{Retrospective studies}

With different target sampling distributions as input, we carried out an extensive retrospective study on 50 slices from the validation set of the FastMRI dataset. The k-space measurements were obtained by applying a forward NUFFT operator ($F$) to the input multi-coil brain MR images. We performed image reconstruction using two different methods:

\subsubsection{CS reconstruction}
\label{sec:cs_recon}
First we used the the synthesis formulation of self-calibrating CS image reconstruction~\cite{ElGueddari_SAM18} by solving for the wavelet coefficients $\mathbf{z}$ as follows:
\begin{equation}
\label{synth_model}
\widehat{\mathbf{z}}=\underset{\mathbf{z} \in \mathbb{C}^{N \times N}}{\operatorname{argmin}} \frac{1}{2} \sum_{\ell=1}^{L}\left\|F_{\Omega} \mathbf{S}_{\ell} \mathbf{\Psi}^* \mathbf{z} - \mathbf{y}_{\ell}\right\|_{2}^{2}+\lambda\|\mathbf{z}\|_{1}
\end{equation}
\noindent where the $L$ is the number of coils. Here the data consistency is enforced with SENSE operators $(F_{\Omega} \mathbf{S}_{\ell})_\ell$, where $F_{\Omega}$ is the NUFFT masked to $\Omega$ and $\mathbf{S}_{\ell}$ is sensitivity map for $\ell^{th}$ coil estimated by density compensated adjoint of the 20\% of acquired k-space center~(see details in~\cite{ElGueddari_SAM18}). $\lambda> 0 $ is the regularization parameter for $\ell_1$-sparsity which was promoted in the wavelet domain $\mathbf{\Psi}$. For our reconstructions, we used Symlet 8 wavelet with 4 scales for $\mathbf{\Psi}$. The regularization parameter $\lambda$ was grid searched between $(10^{-4}, 10^{0})$ while maximizing for the reconstruction quality using structural similarity index (SSIM) in retrospective reconstruction. In order to accelerate convergence, we preconditioned the k-space using density compensation. The compensation weights were estimated with 10 iterations of method as described in~\cite{pipe_dc}. Final MR images were reconstructed as $\widehat{\mathbf{x}}=\mathbf{\Psi}\widehat{\mathbf{z}}$.

\subsubsection{DL reconstruction network (NC-PDNet)}
\label{sec:ncpdnet}
\begin{figure}[h!]
	\begin{center}
		\begin{tabular}{c@{\hspace*{1mm}}c@{\hspace*{1mm}}c@{\hspace*{1mm}}c} 
			&&{\bf (A)} Densities & {\bf (B)} Trajectories
			\\
			\rotatebox[origin=c]{90}{{\bf (i)} VDS}&
			\rotatebox[origin=c]{90}{$\boldsymbol{\rm \rho}_{vds}$ }&
			\parbox[m]{.35\linewidth}{\includegraphics[trim={3.2cm 1cm 2.7cm 1.5cm},clip, width=\linewidth]{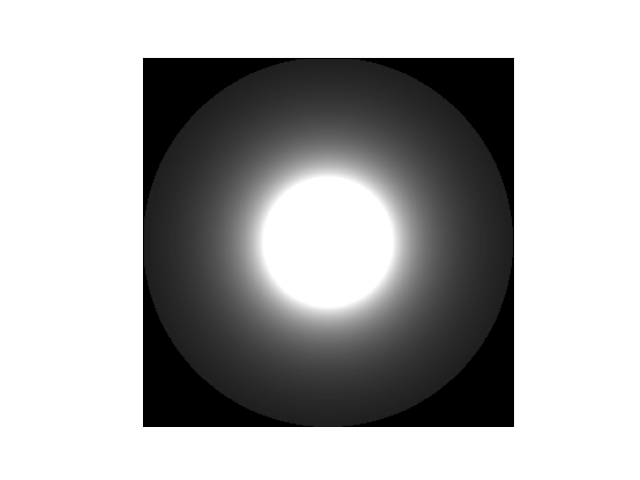}}&
			\parbox[m]{.35\linewidth}{\includegraphics[trim={0.5cm 0.5cm 0.5cm 1.2cm},clip, width=\linewidth]{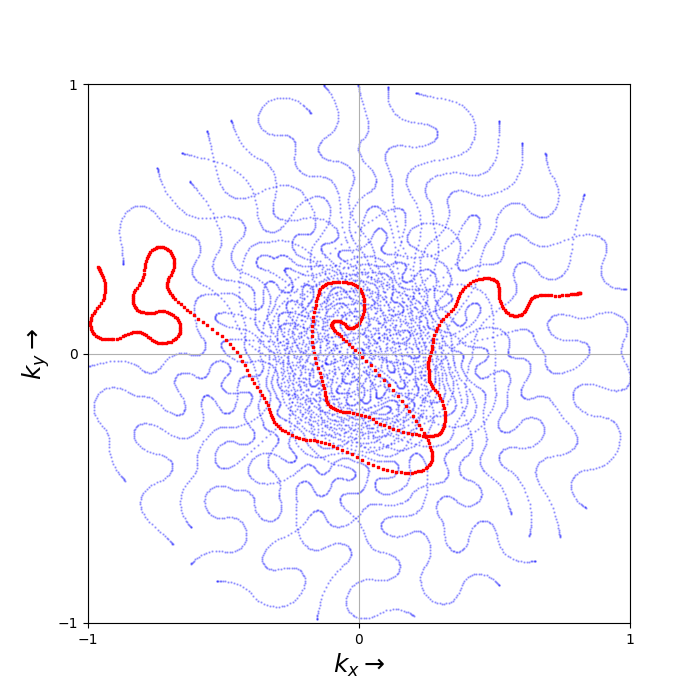}}\\
			\rotatebox[origin=c]{90}{{\bf (ii)} Spectrum}&
			\rotatebox[origin=c]{90}{$\boldsymbol{\rm \rho}_{sb}$ }&
			\parbox[m]{.35\linewidth}{\includegraphics[trim={3.2cm 1cm 2.7cm 1.5cm},clip, width=\linewidth]{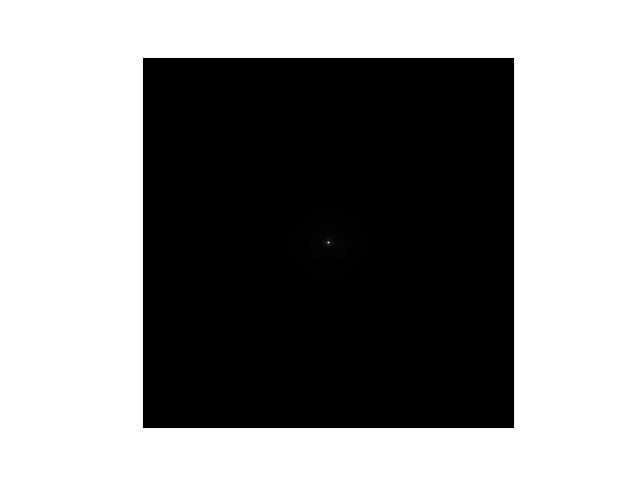}}&
			\parbox[m]{.35\linewidth}{\includegraphics[trim={0.5cm 0.5cm 0.5cm 1.2cm},clip,width=\linewidth]{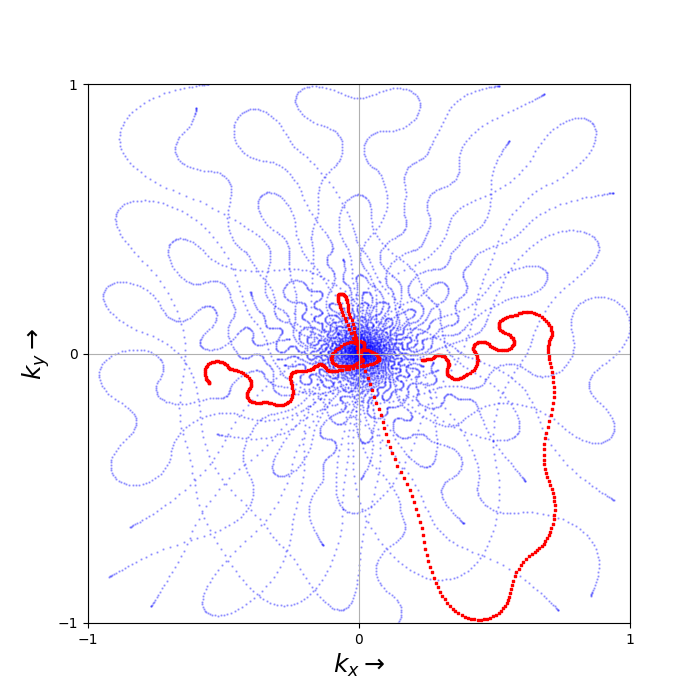}}\\
			\rotatebox[origin=c]{90}{{\bf (iii)} Log-Spectrum}&
			\rotatebox[origin=c]{90}{$\boldsymbol{\rm \rho}_{lsb}$ }&
			\parbox[m]{.35\linewidth}{\includegraphics[trim={3.2cm 1cm 2.7cm 1.5cm},clip, width=\linewidth]{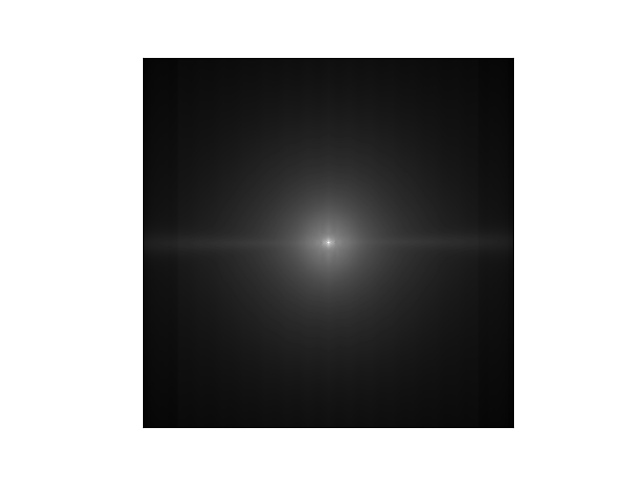}}&
			\parbox[m]{.35\linewidth}{\includegraphics[trim={0.5cm 0.5cm 0.5cm 1.2cm},clip,width=\linewidth]{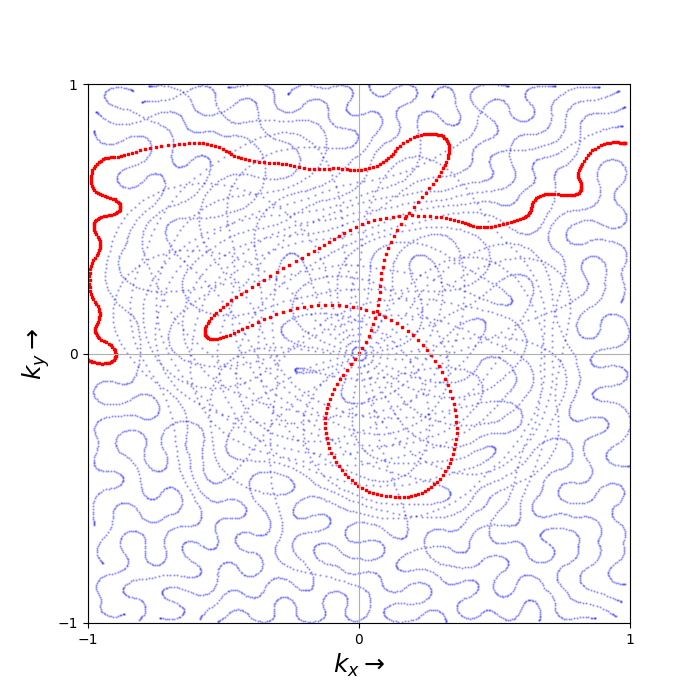}}\\
			\rotatebox[origin=c]{90}{{\bf (iv)} LOUPE}&
			\rotatebox[origin=c]{90}{$\boldsymbol{\rm \rho}_{lb}$ }&
			\parbox[m]{.35\linewidth}{\includegraphics[trim={3.2cm 1cm 2.7cm 1.5cm},clip, width=\linewidth]{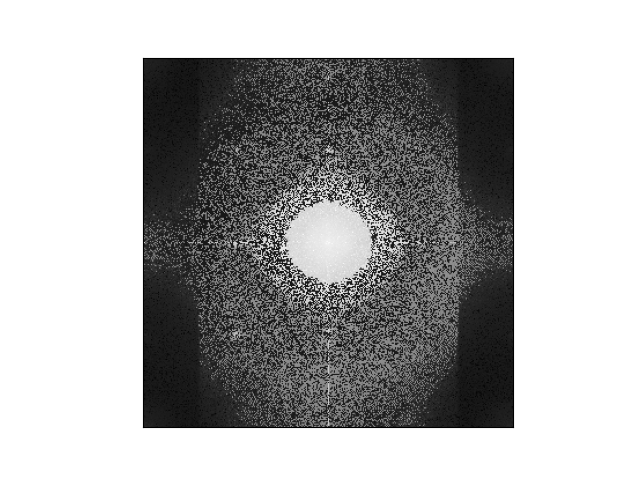}}&
			\parbox[m]{.35\linewidth}{\includegraphics[trim={0.5cm 0.5cm 0.5cm 1.2cm},clip,width=\linewidth]{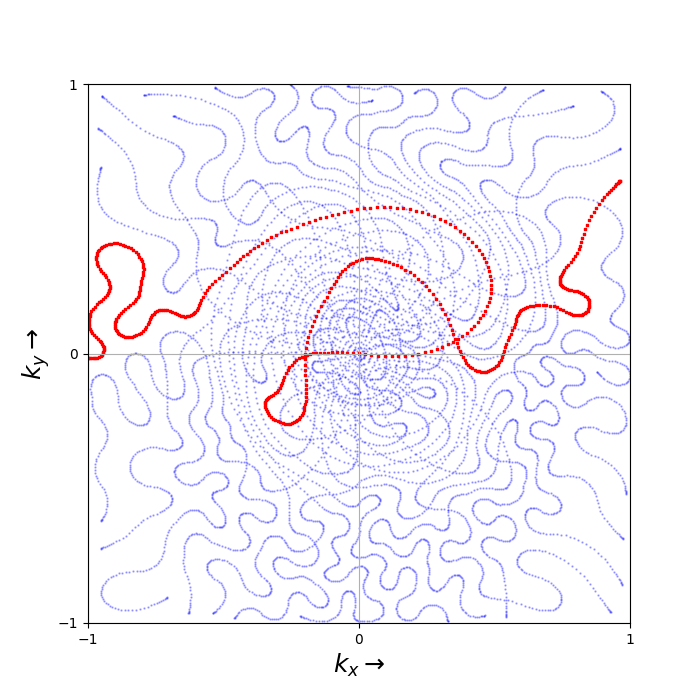}}\\
		\end{tabular}
		\caption{\label{fig:density_n_sparks} {\bf (A):} The target sampling densities obtained for $\text{T}_1$-weighted images with: {\bf (i)} VDS ($\boldsymbol{\rm \rho}_{vds}$), a radially decaying parameterized density, with C=25\% and D=2 in \cite{ChaithyaGR_IEEE_20}; {\bf (ii)} Average spectrum ($\boldsymbol{\rm \rho}_{sb}$) over the dataset based on \cite{Knoll2011}; {\bf (iii)} Average logarithm of the spectrum ($\boldsymbol{\rm \rho}_{lsb}$) over the dataset, to flatten the density in {\bf (ii)}; {\bf (iv)} LOUPE ($\boldsymbol{\rm \rho}_{lb}$) \cite{loupe2020} coupled with XPDNet \cite{ramzi2020xpdnet} reconstruction. {\bf ~(B):} Corresponding k-space trajectories generated with $N_c = 16$ ($R=2.5$), $N_s = 512$, $G_{\rm max} = 40$ mT/m and $S_{\rm max} =180$ T/m/s. For illustration purpose, a single shot is colored in red.}
	\end{center}
\end{figure}

For an extension into DL-based reconstruction, we used NC-PDNet \cite{ramzi2021density}, which is a non-Cartesian extension of the XPDNet used for learning the sampling density. More precisely, we used a density compensated unrolled non-Cartesian reconstruction network, whose parameters are the same as those described in~\cite{ramzi2021density}. This model was trained for 70k gradient descent steps on the respective contrasts ($T_1$-w and $T_2$-w) from multi-coil brain dataset with SPARKLING trajectories obtained in Fig.~\ref{fig:density_n_sparks}. 

\begin{figure*}[htp]
	\begin{center}
		\begin{mdframed}[linewidth=0pt, leftmargin=-1.2cm,rightmargin=-1.2cm, usetwoside=false]
			\resizebox{\linewidth}{!}{
				\begin{tabular}{c@{\hspace*{1mm}}c@{\hspace*{1mm}}|c@{\hspace*{1mm}}c@{\hspace*{1mm}}c} 
					\multicolumn{2}{c}{\bf (A) $T_1$-w Images}&
					\multicolumn{2}{c}{\bf (B) $T_2$-w Images}&
					\\
					{\footnotesize \bf SSIM} & {\footnotesize \bf PSNR} &	{\footnotesize \bf SSIM} & {\footnotesize \bf PSNR} &\\
					\parbox[m]{.25\linewidth}{
						\includegraphics[trim={1cm 1cm 0.4cm 0.7cm},clip, width=\linewidth]{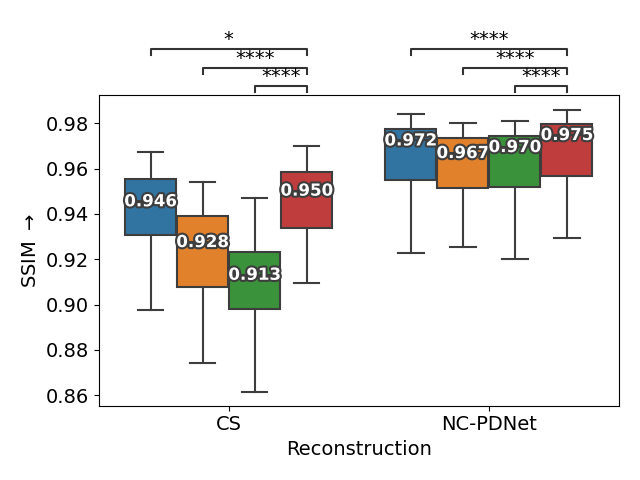}}&
					\parbox[m]{.25\linewidth}{
						\includegraphics[trim={1cm 1cm 0.4cm 0.7cm},clip, width=\linewidth]{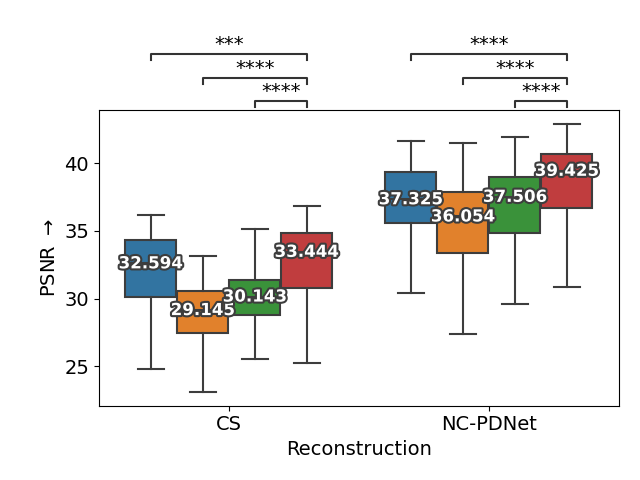}}&
					\parbox[m]{.25\linewidth}{
						\includegraphics[trim={1cm 1cm 0.4cm 0.7cm},clip, width=\linewidth]{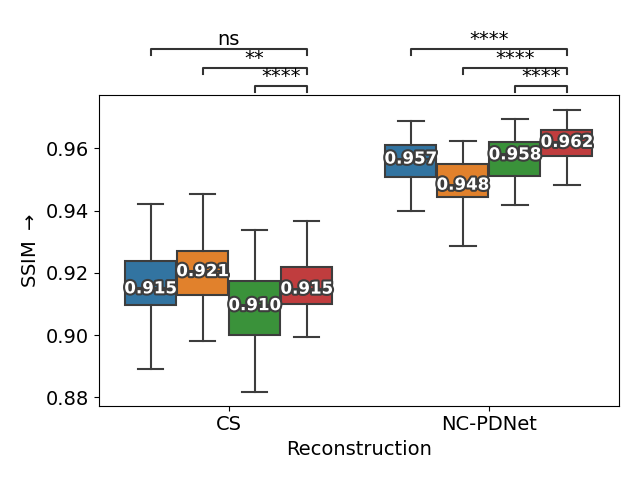}}&
					\parbox[m]{.25\linewidth}{
						\includegraphics[trim={1cm 1cm 0.4cm 0.7cm},clip, width=\linewidth]{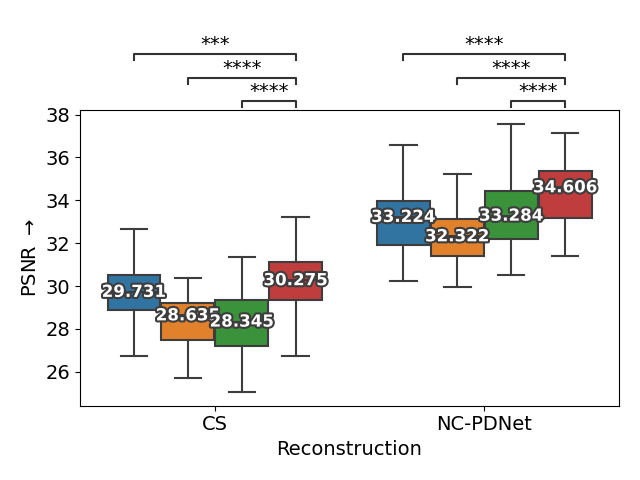}}&
					\parbox[m]{.05\linewidth}{
						\includegraphics[width=\linewidth]{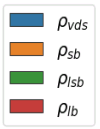}}
			\end{tabular}}
		\end{mdframed}
		\caption{\label{fig:retro_results}Retrospective study on different trajectories for $R=2.5$ on 50 slices of {\bf (A)} $T_1$-w and {\bf (B)} $T_2$-w Images. The reconstructions were performed with both CS based reconstruction (Sec.~\ref{sec:cs_recon}) and using NC-PDNet (Sec.\ref{sec:ncpdnet}) The median SSIM and PSNR scores are indicated.}
	\end{center}
\end{figure*}

\section{Results}
In this section we briefly present the densities and trajectories for various methods of estimating the target sampling densities as described in Sec.~\ref{sec:density_spec}. Then we briefly go through the retrospective reconstruction results that we obtained.
%Later, we proceed to obtain the densities for LOUPE-based method ($\boldsymbol{\rm \rho}_{lb}$) for varying contrasts and acceleration factors.

\subsection{Densities and trajectories}
The varied target sampling densities and their respective SPARKLING trajectories are presented in Fig.~\ref{fig:density_n_sparks}. We see that the direct spectrum-based density $\boldsymbol{\rm \rho}_{sb}$ is extremely dense at the center of k-space, leading to really dense sampling here in the respective trajectories. The {\em log-spectrum} method does indeed flatten out the density $\boldsymbol{\rm \rho}_{lsb}$, allowing the trajectories to explore more high frequencies. Finally, the LOUPE based density does oversample the center of k-space resulting in a scheme very similar to variable density sampling. However, the density $\boldsymbol{\rm \rho}_{lb}$ from LOUPE is more grainy since the learning of this density happens on a Cartesian grid.

\subsection{Retrospective image reconstruction studies}

\subsubsection{Quantitative results}
We carried out retrospective studies on 50 slices of the validation data~(two imaging contrasts, namely $\text{T}_1$ and $\text{T}_2$) in the fastMRI dataset for all the above generated trajectories. 
We computed the SSIM and peak signal-to-noise ratio~(PSNR) metrics on the reconstructed MR images with a mask on the brain in order to assess image quality. We present the results as boxplots and annotated the significance as paired t-test in Fig.~\ref{fig:retro_results}.

Firstly, we note that all methods perform pretty decently as long as the sampling density has been optimized, with NC-PDNet consistently outperforming traditional reconstruction schemes. However, we see that the SPARKLING trajectories with $\boldsymbol{\rm \rho}_{lb}$ densities consistently perform well throughout with SSIMs always larger than 0.95/0.91 for T1/T2 contrast~(red boxes). 
Additionally, this method has the highest PSNR. This confirms our hypothesis that a method which is both data-driven and reconstruction aware outperforms its competitors for learning a target sampling density.
%With respect to SSIM, $\boldsymbol{\rm \rho}_{sb}$ is sometimes slightly better than $\boldsymbol{\rm \rho}_{lb}$.
Finally, we noticed that $\boldsymbol{\rm \rho}_{vds}$ performs similarly to $\boldsymbol{\rm \rho}_{lb}$ with respect to SSIM in most cases. This might be due to the properties of k-space content in brain imaging, which is radially symmetric. Hence optimizing for a radially decaying density gives similar performances to LOUPE-based methods. However, it is worth mentioning that this optimization of parameterized density is very computationally intensive as it involves both trajectory generation and retrospective reconstruction in order to understand which parameter affects the most image quality.
% for multiple parameter sets $(C,D)$ 

\subsubsection{Qualitative results}

\begin{figure*}[h!]
	\begin{center}
		\begin{tabular}{c@{\hspace*{1mm}}c@{\hspace*{1mm}}c@{\hspace*{1mm}}c@{\hspace*{1mm}}c}
			
			\textbf{Reference $T_1$-w Image}&{\bf (i)} $\boldsymbol{\rm \rho}_{vds}$  & {\bf (ii)} $\boldsymbol{\rm \rho}_{sb}$ & {\bf (iii)} $\boldsymbol{\rm \rho}_{lsb}$ & {\bf (iii)} $\boldsymbol{\rm \rho}_{lb}$
			\\
			\parbox[m]{.17\linewidth}{\includegraphics[trim={4.7cm 1cm 4.5cm 0.7cm},clip, width=\linewidth]{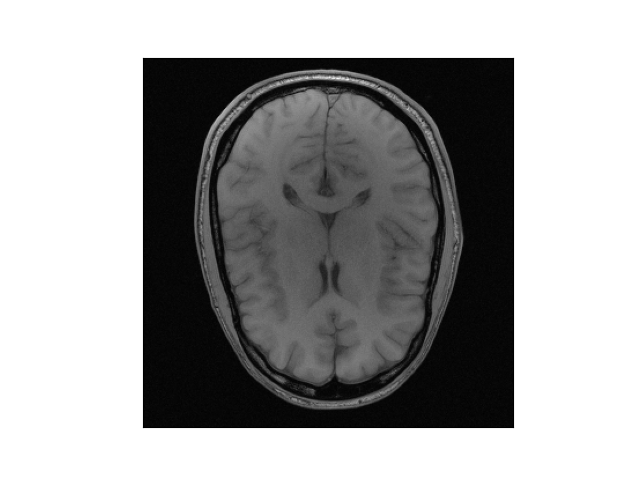}} &
			\parbox[m]{.17\linewidth}{\includegraphics[trim={4.7cm 1cm 4.5cm 0.7cm},clip, width=\linewidth]{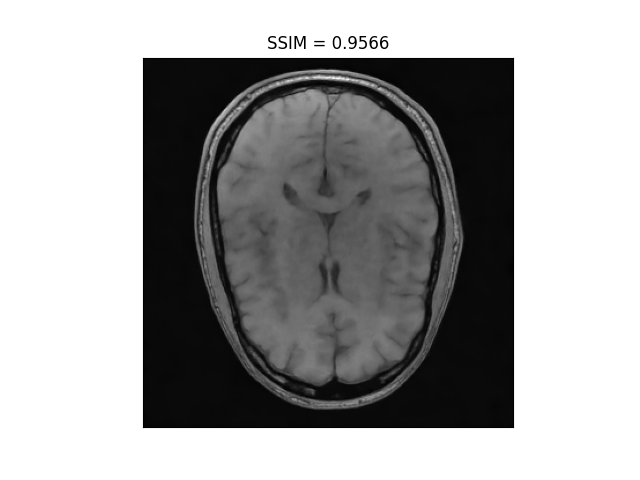}}&
			\parbox[m]{.17\linewidth}{\includegraphics[trim={4.7cm 1cm 4.5cm 0.7cm},clip, width=\linewidth]{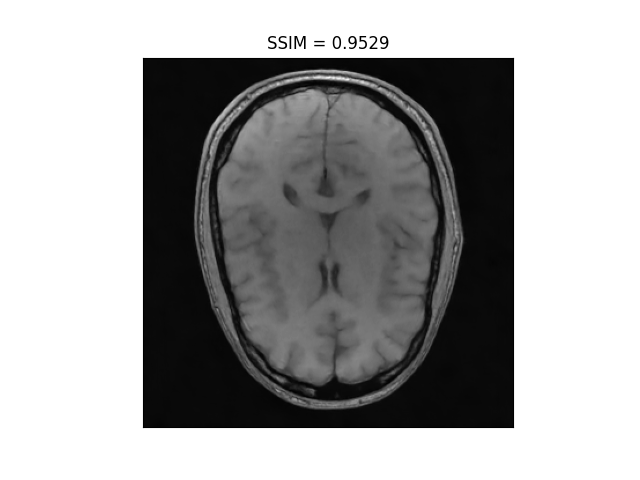}}&
			\parbox[m]{.17\linewidth}{\includegraphics[trim={4.7cm 1cm 4.5cm 0.7cm},clip, width=\linewidth]{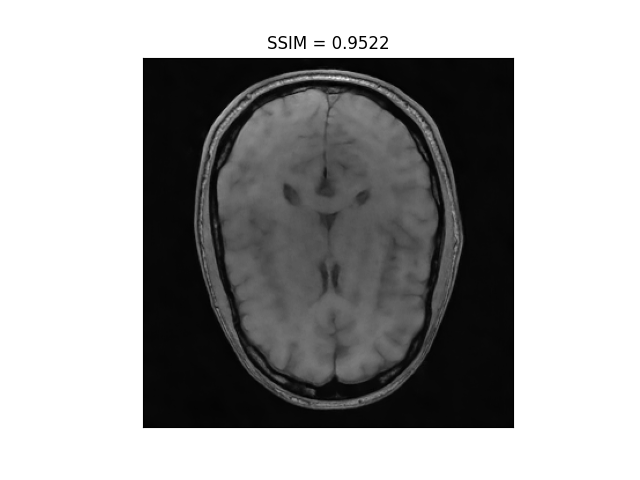}}&
			\parbox[m]{.17\linewidth}{\includegraphics[trim={4.7cm 1cm 4.5cm 0.7cm},clip, width=\linewidth]{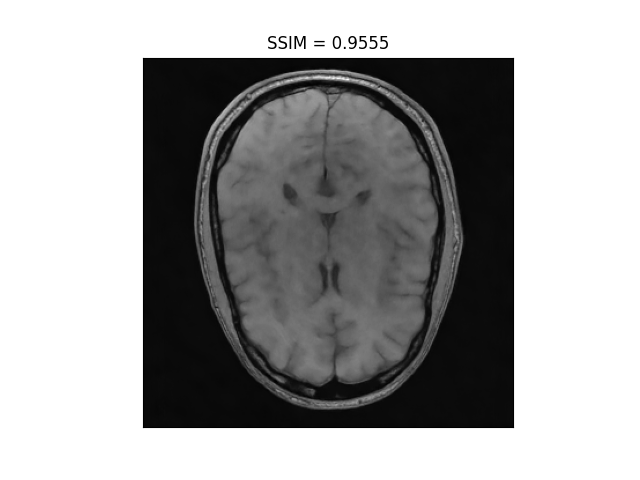}}
			\\
			\comment{&
			\parbox[m]{.17\linewidth}{\includegraphics[trim={4.7cm 1cm 4.5cm 0.7cm},clip, width=\linewidth]{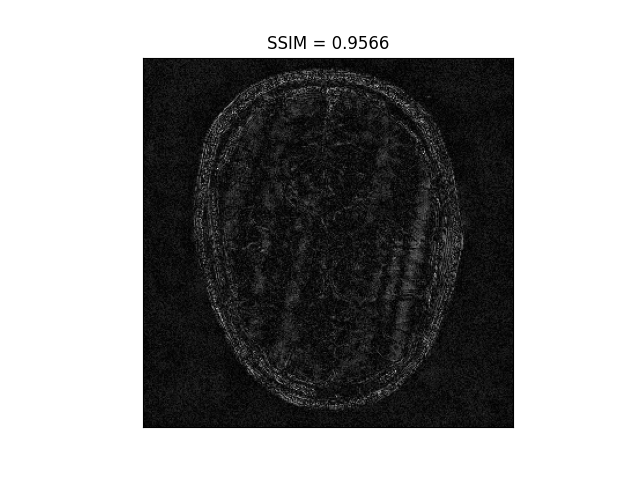}}&
			\parbox[m]{.17\linewidth}{\includegraphics[trim={4.7cm 1cm 4.5cm 0.7cm},clip, width=\linewidth]{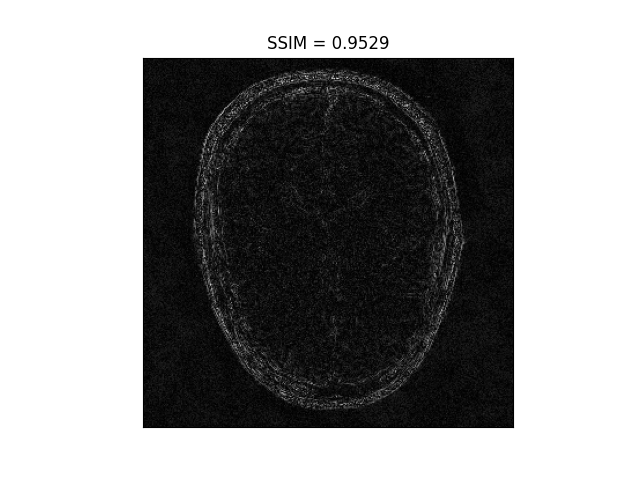}}&
			\parbox[m]{.17\linewidth}{\includegraphics[trim={4.7cm 1cm 4.5cm 0.7cm},clip, width=\linewidth]{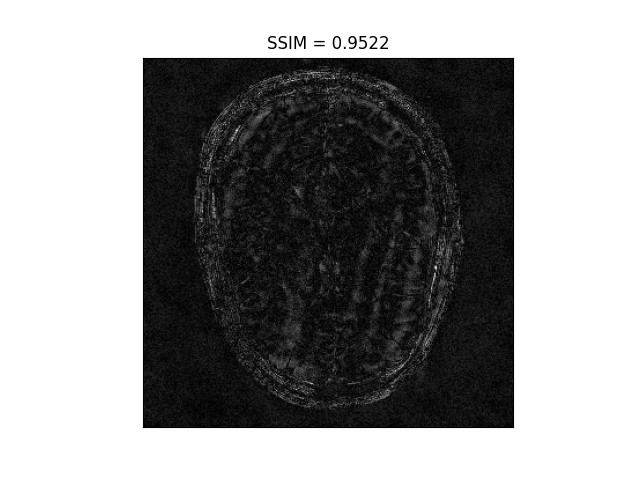}}&
			\parbox[m]{.17\linewidth}{\includegraphics[trim={4.7cm 1cm 4.5cm 0.7cm},clip, width=\linewidth]{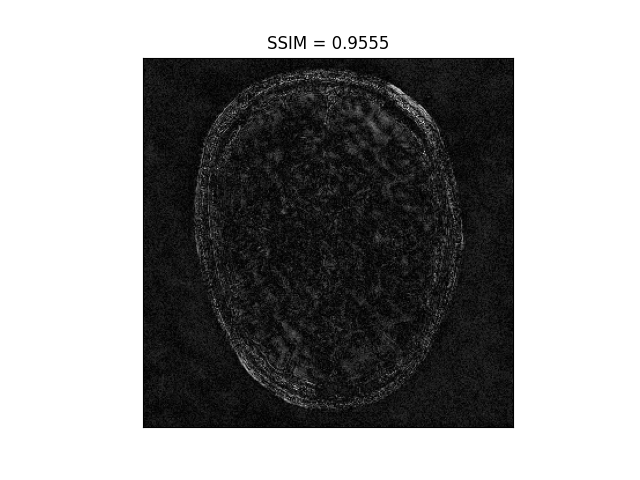}}
		}
		\end{tabular}
		\caption{\label{fig:T1_results}
		NC-PDNet-based image reconstruction for retrospective $T1$-w imaging with slice $\slicex$ in \textit{file\_brain\_AXT1\_201\_6002725.h5} from validation data in fastMRI dataset for different target sampling densities.}
	\end{center}
\end{figure*}

\begin{figure*}[h!]
	\begin{center}
		\begin{tabular}{c@{\hspace*{1mm}}c@{\hspace*{1mm}}c@{\hspace*{1mm}}c@{\hspace*{1mm}}c}
			
			\textbf{Reference $T_2$-w Image}&{\bf (i)} $\boldsymbol{\rm \rho}_{vds}$  & {\bf (ii)} $\boldsymbol{\rm \rho}_{sb}$ & {\bf (iii)} $\boldsymbol{\rm \rho}_{lsb}$ & {\bf (iii)} $\boldsymbol{\rm \rho}_{lb}$
			\\
			\parbox[m]{.17\linewidth}{\includegraphics[trim={4.3cm 1cm 4.1cm 0.7cm},clip, width=\linewidth]{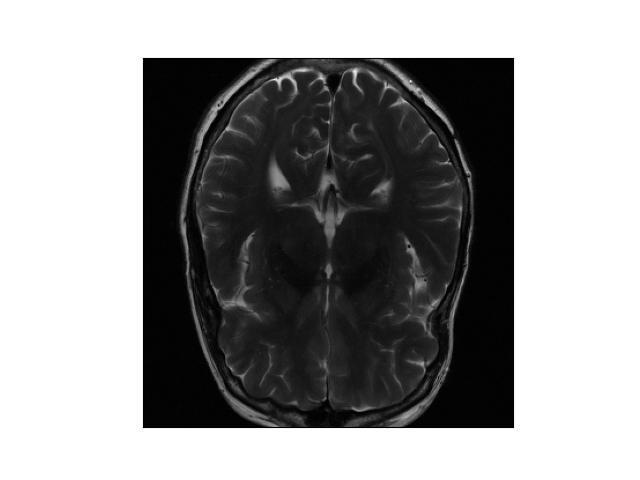}}	&
			\parbox[m]{.17\linewidth}{\includegraphics[trim={4.3cm 1cm 4.1cm 0.7cm},clip, width=\linewidth]{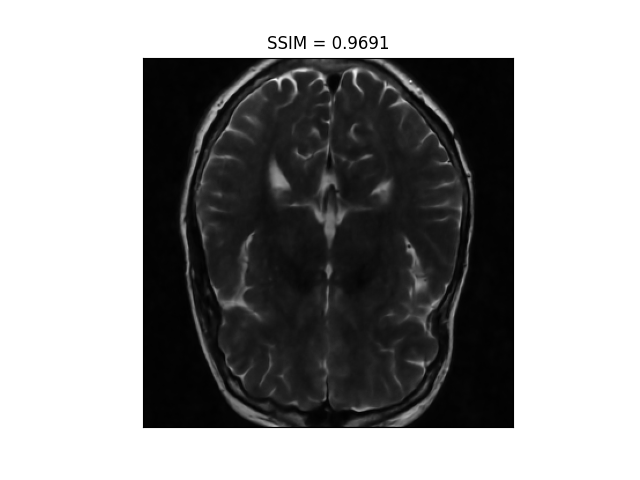}}&
			\parbox[m]{.17\linewidth}{\includegraphics[trim={4.3cm 1cm 4.1cm 0.7cm},clip, width=\linewidth]{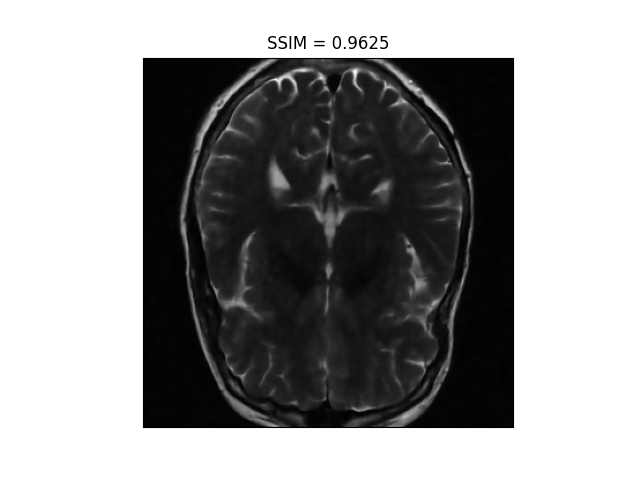}}&
			\parbox[m]{.17\linewidth}{\includegraphics[trim={4.3cm 1cm 4.1cm 0.7cm},clip, width=\linewidth]{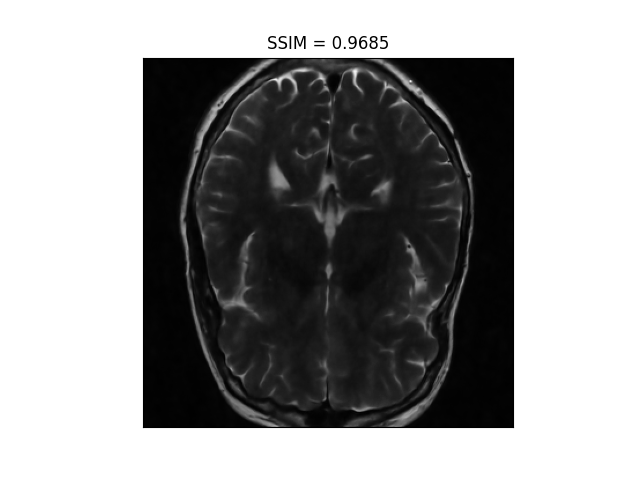}}&
			\parbox[m]{.17\linewidth}{\includegraphics[trim={4.3cm 1cm 4.1cm 0.7cm},clip, width=\linewidth]{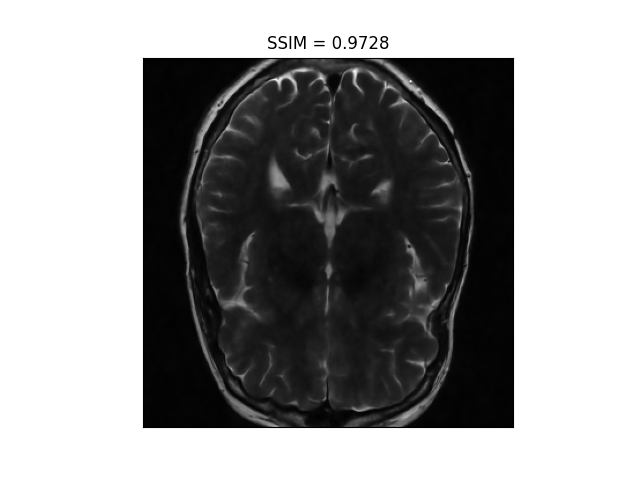}}
			\\	
			\comment{
			&
			\parbox[m]{.17\linewidth}{\includegraphics[trim={4.3cm 1cm 4.1cm 0.7cm},clip, width=\linewidth]{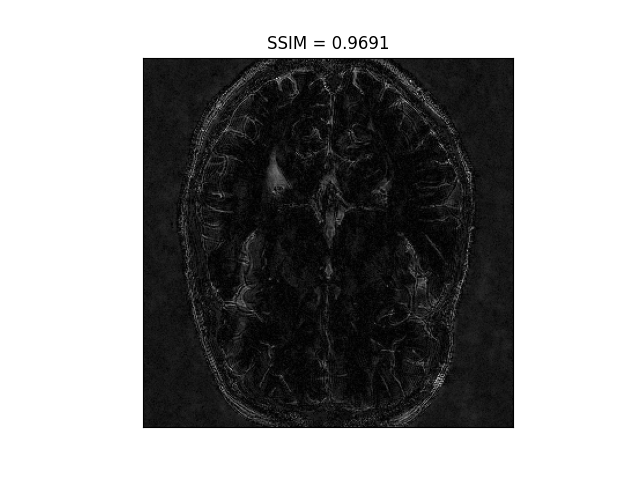}}&
			\parbox[m]{.17\linewidth}{\includegraphics[trim={4.3cm 1cm 4.1cm 0.7cm},clip, width=\linewidth]{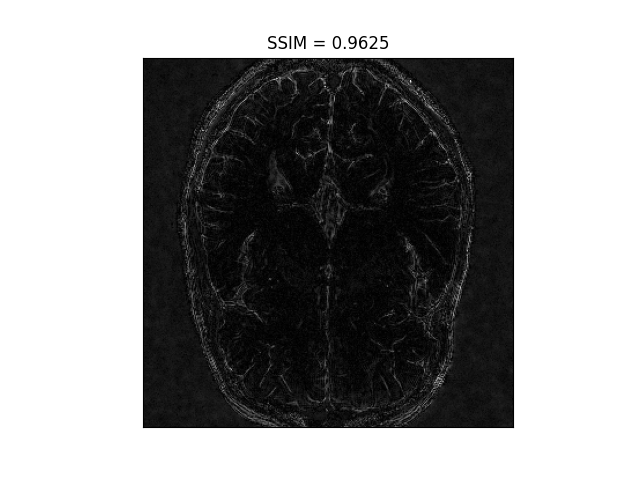}}&
			\parbox[m]{.17\linewidth}{\includegraphics[trim={4.3cm 1cm 4.1cm 0.7cm},clip, width=\linewidth]{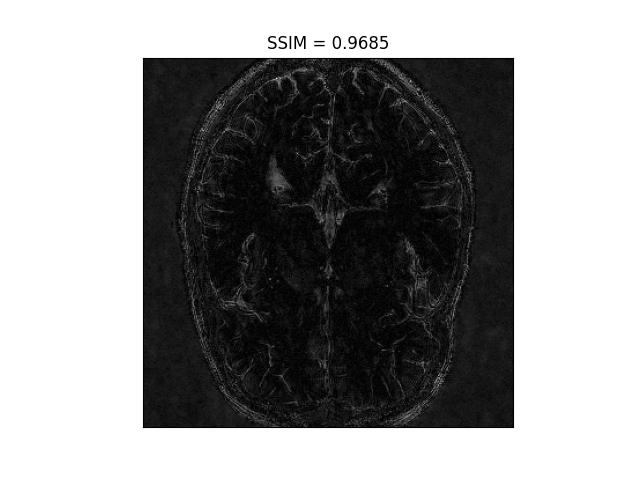}}&
			\parbox[m]{.17\linewidth}{\includegraphics[trim={4.3cm 1cm 4.1cm 0.7cm},clip, width=\linewidth]{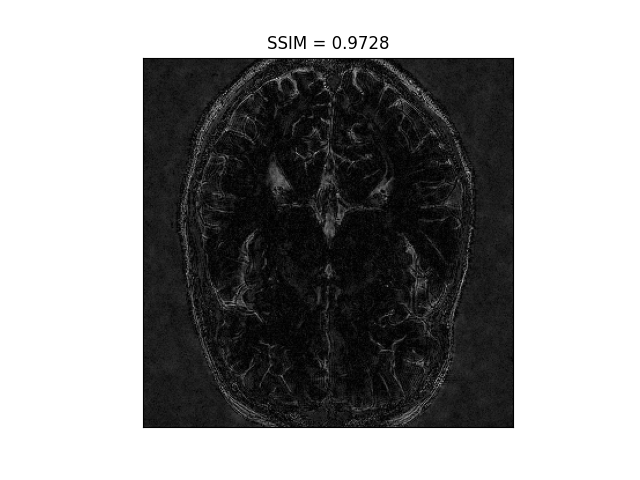}}
		}
		\end{tabular}
		\caption{\label{fig:T2_results}
		 NC-PDNet-based image reconstruction for retrospective $T2$-w imaging with slice $\slicey$ in \textit{file\_brain\_AXT2\_200\_2000019.h5} from validation data in fastMRI dataset for different target sampling densities.}
	\end{center}
\end{figure*}

For visual inspection, we present the results of image reconstruction from data undersampled using SPARKLING trajectories generated for various target densities in Fig.~\ref{fig:T1_results}~($\text{T}_1$-w images) and Fig.~\ref{fig:T2_results}~($\text{T}_2$-w images). For the sake of space, we only report the best reconstruction results, i.e. with NC-PDNet.
For $\text{T}_1$-weighted contrast, we show that all methods give similarly performing results, however $\boldsymbol{\rm \rho}_{vds}$ and $\boldsymbol{\rm \rho}_{lb}$  provide the best SSIM scores. Further, we observe that in this case, $\boldsymbol{\rm \rho}_{vds}$ is slightly better than $\boldsymbol{\rm \rho}_{lb}$. On the contrary, for $\text{T}_2$-w contrast, $\boldsymbol{\rm \rho}_{lb}$ outperforms the other densities as reflected both visually in Fig.~\ref{fig:T2_results} and quantitatively~(see Fig.~\ref{fig:retro_results}).

\section{Conclusions}
In this study, we addressed the main drawback of the SPARKLING algorithm, namely the need for a good target sampling density as an input parameter. We setup four different methods to generate optimized target sampling densities and design SPARKLING trajectories accordingly. We showed that the LOUPE-based approach is the most promising as it provides consistent results across contrasts. 
%We see all the methods except log-spectrum based density perform pretty well, and LOUPE based method is consistent across contrasts, as it jointly optimizes for sampling density and Cartesian reconstruction network (XPDNet). 
A limitation of this work is that there remains some
split between the acquisition and reconstruction models in a fully non-Cartesian setting.
Under the current study, the sampling density was jointly optimized with a Cartesian DL reconstruction network. Then, non-Cartesian SPARKLING trajectories were generated and retrospective validation was performed using a non-Cartesian DL network. There is thus still a gap between the training and validation stage in this pipeline.
%Due to this disjointedness, our reconstruction model could not have adapted well to the non-Cartesian trajectories with learned densities under a Cartesian reconstruction framework.
In spite of this limitation, we obtained promising results. In terms of perspective, we plan to work on a joint network between NC-PDNet and SPARKLING to efficiently learn the k-space trajectories in a data-driven manner, under the MR Hardware constraints.

%Further, this study does not model the SNR and noise in the images, which could be a key added factor for prospective studies. We plan to integrate SNR model based on the scanner constraints to have a retrospective study which can better model the real world in our future works.

\section*{Acknowledgments}
We acknowledge the French Institute of development and resources in scientific computing for their AI program allowing us to use the Jean Zay supercomputer's GPU partitions.

\bibliographystyle{IEEEtran}
\bibliography{ref.bib}

\begin{thebibliography}{10}
\providecommand{\url}[1]{#1}
\csname url@samestyle\endcsname
\providecommand{\newblock}{\relax}
\providecommand{\bibinfo}[2]{#2}
\providecommand{\BIBentrySTDinterwordspacing}{\spaceskip=0pt\relax}
\providecommand{\BIBentryALTinterwordstretchfactor}{4}
\providecommand{\BIBentryALTinterwordspacing}{\spaceskip=\fontdimen2\font plus
\BIBentryALTinterwordstretchfactor\fontdimen3\font minus
  \fontdimen4\font\relax}
\providecommand{\BIBforeignlanguage}[2]{{%
\expandafter\ifx\csname l@#1\endcsname\relax
\typeout{** WARNING: IEEEtran.bst: No hyphenation pattern has been}%
\typeout{** loaded for the language `#1'. Using the pattern for}%
\typeout{** the default language instead.}%
\else
\language=\csname l@#1\endcsname
\fi
#2}}
\providecommand{\BIBdecl}{\relax}
\BIBdecl

\bibitem{Lustig2007}
M.~Lustig, D.~Donoho, and J.~M. Pauly, ``Sparse {MRI}: The application of
  compressed sensing for rapid {MR} imaging,'' \emph{{{M}agn. {R}eson.
  {M}ed.}}, vol.~58, no.~6, pp. 1182--1195, 2007.

\bibitem{puy2011variable}
G.~Puy, P.~Vandergheynst, and Y.~Wiaux, ``On variable density compressive
  sampling,'' \emph{{{IEEE} {S}ignal {P}rocess. {L}ett.}}, vol.~18, no.~10, pp.
  595--598, 2011.

\bibitem{chauffert2013variable}
N.~Chauffert, P.~Ciuciu, and P.~Weiss, ``Variable density compressed sensing in
  mri. theoretical vs heuristic sampling strategies,'' in \emph{2013 IEEE 10th
  International Symposium on Biomedical Imaging}.\hskip 1em plus 0.5em minus
  0.4em\relax IEEE, 2013, pp. 298--301.

\bibitem{Chauffert_SIAM2014}
N.~Chauffert \emph{et~al.}, ``Variable density sampling with continuous
  trajectories. {A}pplication to {MRI},'' \emph{{{SIAM} {J}. {I}mag. {S}ci.}},
  vol.~7, no.~4, pp. 1962--1992, Nov. 2014.

\bibitem{adcock2017breaking}
B.~Adcock \emph{et~al.}, ``Breaking the coherence barrier: A new theory for
  compressed sensing,'' in \emph{Forum of Mathematics, Sigma}, vol.~5.\hskip
  1em plus 0.5em minus 0.4em\relax Cambridge University Press, 2017.

\bibitem{boyer2017compressed}
C.~Boyer, J.~Bigot, and P.~Weiss, ``Compressed sensing with structured sparsity
  and structured acquisition,'' \emph{{{A}ppl. {C}omput. {H}armon. {A}nal.}},
  vol.~46, no.~2, pp. 312--350, 2019.

\bibitem{radial_spokes}
P.~C. Lauterbur, ``Image formation by induced local interactions: Examples
  employing nuclear magnetic resonance,'' \emph{Nature}, vol. 242, no. 5394,
  pp. 190--191, Mar 1973.

\bibitem{ahn1986high}
C.~Ahn \emph{et~al.}, ``High-speed spiral-scan echo planar {NMR imaging-I},''
  \emph{{{IEEE} {T}rans. {M}ed. {I}mag.}}, vol.~5, no.~1, pp. 2--7, 1986.

\bibitem{Lazarus_MRM_19}
C.~Lazarus \emph{et~al.}, ``{SPARKLING: variable-density k-space filling curves
  for accelerated T2*-weighted MRI},'' \emph{{{M}agn. {R}eson. {M}ed.}},
  vol.~81, no.~6, pp. 3643--3661, 2019.

\bibitem{Lazarus_NMRB_20}
------, ``{3D variable-density SPARKLING trajectories for high-resolution
  T2*-weighted magnetic resonance imaging},'' \emph{{{NMR} {B}iomed.}}, 2020.

\bibitem{ChaithyaGR_IEEE_20}
G.~R. Chaithya \emph{et~al.}, ``{Globally optimized 3D SPARKLING trajectories
  for high-resolution T2*-weighted Magnetic Resonance imaging},'' \emph{under
  review} at \emph{IEEE Transactions on Medical Imaging}, May 2021.

\bibitem{Knoll2011}
F.~Knoll \emph{et~al.}, ``Adapted random sampling patterns for accelerated
  {MRI},'' \emph{MRM in Physics, Biology and Medicine}, vol.~24, no.~1, pp.
  43--50, Feb 2011.

\bibitem{Knoll_2020}
------, ``{Advancing machine learning for MR image reconstruction with an open
  competition: Overview of the 2019 fastMRI challenge},'' \emph{{MRM}},
  vol.~84, no.~6, p. 3054–3070, Jun 2020.

\bibitem{muckley2021results}
M.~J. Muckley \emph{et~al.}, ``{Results of the 2020 fastMRI Challenge for
  Machine Learning MR Image Reconstruction},'' \emph{IEEE Transactions on
  Medical Imaging}, 2021.

\bibitem{ramzi_bench2020}
Z.~Ramzi, P.~Ciuciu, and J.-L. Starck, ``{Benchmarking Deep Nets MRI
  Reconstruction Models on the FastMRI Publicly Available Dataset},'' in
  \emph{2020 IEEE 17th ISBI}, Iowa City (virtual), United States, Apr. 2020,
  pp. 1441--1445.

\bibitem{sherry2020learning}
F.~Sherry \emph{et~al.}, ``{Learning the sampling pattern for MRI},''
  \emph{{{IEEE} {T}rans. {M}ed. {I}mag.}}, 2020.

\bibitem{loupe2020}
C.~D. Bahadir \emph{et~al.}, ``{Deep-learning-based optimization of the
  under-sampling pattern in MRI},'' \emph{{{IEEE} {T}rans. {C}omput.
  {I}maging}}, vol.~6, pp. 1139--1152, 2020.

\bibitem{weiss2019pilot}
T.~Weiss \emph{et~al.}, ``{PILOT: Physics-informed learned optimal trajectories
  for accelerated MRI},'' \emph{arXiv:1909.05773v5}, Aug. 2020.

\bibitem{vedula20203d}
S.~Vedula \emph{et~al.}, ``{3D FLAT: Feasible Learned Acquisition Trajectories
  for Accelerated MRI},'' in \emph{3rd Intern. WS MLMIR 2020, Held in
  Conjunction with MICCAI 2020}.\hskip 1em plus 0.5em minus 0.4em\relax Lima,
  Peru: Springer Nature, Oct. 2020.

\bibitem{ramzi2021density}
Z.~Ramzi, J.-L. Starck, and P.~Ciuciu, ``{Density Compensated Unrolled Networks
  for Non-Cartesian MRI Reconstruction},'' in \emph{2021 IEEE 18th ISBI}, Nice,
  France (virtual), pp. 1443--1447.

\bibitem{chauffert2017projection}
N.~Chauffert \emph{et~al.}, ``A projection method on measures sets,''
  \emph{{{C}onstr. {A}pprox.}}, vol.~45, no.~1, pp. 83--111, 2017.

\bibitem{zbontar2019fastmri}
J.~Zbontar \emph{et~al.}, ``{fastMRI: An Open Dataset and Benchmarks for
  Accelerated MRI},'' 2019.

\bibitem{unet}
D.~{Lee}, J.~{Yoo}, and J.~C. {Ye}, ``Deep residual learning for compressed
  sensing {MRI},'' in \emph{ISBI 2017}, pp. 15--18.

\bibitem{ramzi2020xpdnet}
Z.~Ramzi, J.-L. Starck, and P.~Ciuciu, ``{XPDNet for MRI Reconstruction: An
  application to the 2020 fastMRI challenge},'' in \emph{2021 ISMRM annual
  meeting}, no. Abstract 275, May 2021.

\bibitem{ElGueddari_SAM18}
E.~Gueddari \emph{et~al.}, ``Self-calibrating nonlinear reconstruction
  algorithms for variable density sampling and parallel reception {MRI},'' in
  \emph{10th {IEEE SAM Signal Processing WS}}, Sheffield, UK, Jul. 2018, pp.
  415--419.

\bibitem{pipe_dc}
J.~G. Pipe \emph{et~al.}, ``Sampling density compensation in {MRI},''
  \emph{{{M}agn. {R}eson. {M}ed.}}, vol.~41, no.~1, pp. 179--186, 1999.

\end{thebibliography}

\end{document}